\documentclass[12pt]{article}
\usepackage{amsmath,amsfonts,amssymb}
\usepackage{graphicx}
\usepackage{supertabular}
\usepackage{fancyhdr}
\usepackage{placeins}
\usepackage{enumitem}
\usepackage{booktabs}
\begin{document}

\title{Likelihood Models for Forensic Genealogy}

\author{William H.~Press\\
  Departments of Computer Science and Integrative Biology\\
  The University of Texas at Austin
\and
  John Hawkins\\
  Oden Institute of Computational Engineering and Science\\
  The University of Texas at Austin}
\maketitle

\newcommand{\onebf}{\boldsymbol{1}}
\newcommand{\Exp}{\text{Exp}}
\newcommand{\Gam}{\text{Gamma}}
\newcommand{\Poi}{\text{Poi}}
\newcommand{\Sigmabf}{\boldsymbol{\Sigma}}
\newcommand{\mubf}{\boldsymbol{\mu}}
\newcommand{\prob}{\text{prob}}
\newcommand{\diag}{\text{diag}}
\newcommand{\mystrut}{\vphantom{\Bigg)}}
\newcommand{\Prev}{P_\text{rev}}
\newcommand{\nbar}{\overline{n}}
\newcommand{\z}{\mathbb{Z}}
\newcommand{\Indent}{\hspace{24 pt}}
\newcommand\bigforall{\mbox{\Large $\mathsurround0pt\forall$}} 
\newcommand{\half}{\tfrac{1}{2}}
\newcommand{\timeseq}{\mathrel{*}=}
\newcommand{\Sbf}{\mathbf{S}}
\newcommand{\Vbf}{\mathbf{V}}
\newcommand{\calM}{{\cal M}}
\thispagestyle{fancy}

\begin{abstract}
  In the idealized Morgan model of crossover, we study the probability
  distributions of shared DNA (identical by descent) between
  individuals having a wide range of relationships (not just lineal
  descendants), especially cases for which previous work produces
  inaccurate results.  Using Monte Carlo simulation, we show that a
  particular, complicated functional form with just one continuous
  fitted parameter accurately approximates the distributions in all
  cases tried.  Analysis of that functional form shows that it is
  close to a normal distribution, not in shared fraction $f$, but in the
  square-root of $f$.  We describe a multivariate normal model in this
  variable for use as a practical framework for several general tasks in
  forensic genealogy that are currently done by less-accurate and less
  well-founded methods.
\end{abstract}

\section{Introduction}\label{intro}

Forensic genealogy seeks the identity of an unknown person (UP) for
whom a DNA sample is available, but whose identified genome is not in
any database.  The UP may be an at-large criminal suspect whose DNA is
present at a crime scene \cite{gskiller}, or he/she may be an
unidentified, deceased victim \cite{babydna}. Alternatively, UPs may be
adopted individuals seeking to know their biological parents, in
which case they are ``unknown'' in pedigree, not name.

The current methodology of forensic genealogy is to genotype or
partially sequence the UP's DNA using a commercial SNP microarray, and
then to upload the file of SNP results in one of several standard
formats to a public genome database such as GEDmatch \cite{GEDmatch}.
Remotely on that database, a similarity match against all available
genomes is performed, returning a list of genomes that are
statistically significant matches to UP, along with identifying
information that the owners of those matches have authorized for
public release.  The similarity score generally reported is the
centimorgan length of autosomal matches, exact or near-exact, of the
diploid genomes, approximately equivalent to the fraction of autosomal
genome that is identical by descent (IBD) \cite{ancestrymatching}.
With current public database sizes, it is common that half a dozen or
more matches to UPs are reported, with centimorgan scores that may be
typical of third or fourth cousins, but closer relatives in favorable
cases.

The job of the forensic genealogist is next to construct a probable
family tree (or pedigree) from the available match data, and then to
place UP convincingly in that tree.  Often, matched individuals may
not even know that they are mutually related, or exactly how.  The
genealogist makes use of public records, commonality of names and
geographical locations, and other data to hypothesize possible family
trees and hypothetical positions for UP in those trees.  This
conventional, often ingenious, detective work frequently yields more
than one seemingly viable hypothesis. Some kind of quantitative
calculation of the relative likelihood of the various hypotheses,
given the observed centimorgan scores, is thus required.

If fractional genomic matches were deterministic (e.g., if the genomes
of third cousins were {\em exactly} 1/128 identical by descent), then
a probabilistic calculation might be unnecessary.  Since, however, the
number of crossovers on each chromosome in each meiosis event is
stochastic, the centimorgan similarity scores of even fairly close
relatives can be intrinsically ambiguous as to what relationship is
implied.  Current practice is to compare the observed centimorgan
scores to distributions derived from the self-reported relationships
of users who upload their genomes and link themselves to other
uploaded genomes \cite{ancestrymatching}. Each centimorgan score is
then assigned a probability for its being at a distance of $k$
meioses, for $k=1,2,3,\ldots$, with the probabilities for all $k$
summing to unity.  In essence, this is a Bayesian calculation on the
self-reported empirical data with a uniform prior on $k$.  The
probabilities for the UP's multiple matches are then most commonly
combined by multiplying the individual probabilities as if they were
independent \cite{wato}. Hypothesized family trees and the location of
UP on them can then be assigned relative probabilities proportional to
these products, again a kind of Bayes but now with a uniform prior on
the hypotheses.

There are various large and small ways that one might improve on the
above procedures.  That is the subject of this paper.  First, one
might want to use probability distributions for the fractional IBD
match of various familial relationships that come from genetics, not
self-reported data.  Self-reporting on distant cousin relationships
may, for example, be subject to error, as may be the self-reporting of
complicated family trees.  Second, one might want not to combine
distinct relationships into $k$-meioses classes.  For example, the IBD
distributions for grandchild and nephew are different, even though
their means of 1/4 shared fraction are the same. (We will show this
below.) Third, one might want to take into account that the
probabilities for multiple matches to the UP are not independent.  For
example, if UP's match to individual P is, by chance, higher than
average, then UP's match to some Q who is a child of P will also tend
to be quite significantly higher than average. (We will make this
quantitative below.) Fourth, the priors assumed by the naive
methodology can be inconsistent.  For example, the uniform prior on
$k$, the meioses distance, may include significant probabilities for
values of $k$ that occur in no allowed (by other data) family tree,
distorting the results for the allowed family trees.

Sections \ref{previouswork} and \ref{results} of this paper address
the first two issues above, as regards the ``classical''
calculation of the distribution of fraction of shared autosomal genome
between related individuals.  We will see that previous calculations,
in work going back to the 1950s, yield unreliable answers.  We will
give analytical formulas that, with a small number of fitted
parameters (whose values we give), produce accurate estimates.  Then,
\S\ref{mvariate} of this paper addresses issues three and four
above, that is, how to construct consistent statistical models that
can compare whole family trees given multiple measured centimorgan
values, accounting for correlations among the measurements.

For mathematical definiteness, we throughout this paper assume a pure
Morgan model, in which the probability of a crossover in meiosis is a
Poisson random process along the length of each chromosome, with rate
1\% probability per centimorgan (by the definition of genetic
distance).  It is known, of course, that this is only an
approximation, and that the distribution of crossovers is
evolutionarily regulated with both signs, that is with both obligate
crossover and crossover interference being observed (see, e.g.,
\cite{jones}, \cite{mezard}, and references therein). While these are
biologically significant effects, we believe the idealizations of
genetic distance and random crossover to be adequate for the purposes
of this paper, and in any case, given current practice, forward
progress for forensic genealogy.

\section{Previous Work}
\label{previouswork}
\subsection{Assuming Poisson Numbers of Common Segments}

Although the work of Morgan, famously confirmed experimentally by
McLintock and others, could have allowed some estimate of the
distribution of fractional shared genome as early as the 1930s, the
first such calculations seem to be those of Fisher \cite{fuller} and
Bennett \cite{bennett} in the 1950s.  These authors noted that, for a
chromosome of genetic length $L$ (in Morgans), there would be in the
mean $kL$ crossover points after $k$ meioses.  On the other hand, the
fraction of common genome with a specific ancestor must be, on
average, $2^{-k}$, by symmetry among $2^k$ ancestors.  These facts,
plus an assumption that the number of common segments is Poisson
distributed, yielded the analytic result
\begin{equation}
  p(f|k,L)\,df = \exp[-k L(2^{-k}+2f)]\,kL\sqrt{2^{-k+1}/f}\,
  I_1\left(2kL\sqrt{2^{-k+1}f}\right)\,df
\label{eqpdf1}
\end{equation}
where $p(f|k,L)$ is the probability density for a fraction $f$ in
common, with $0<f\le 1/2$.  There is also a massed probability for
no commonality ($f=0$) given by
\begin{equation}
p(0|k,L) = \exp(-k 2^{-k} L)
\end{equation}
\begin{figure}[!tb]
\centering
\includegraphics[width=5.5in]{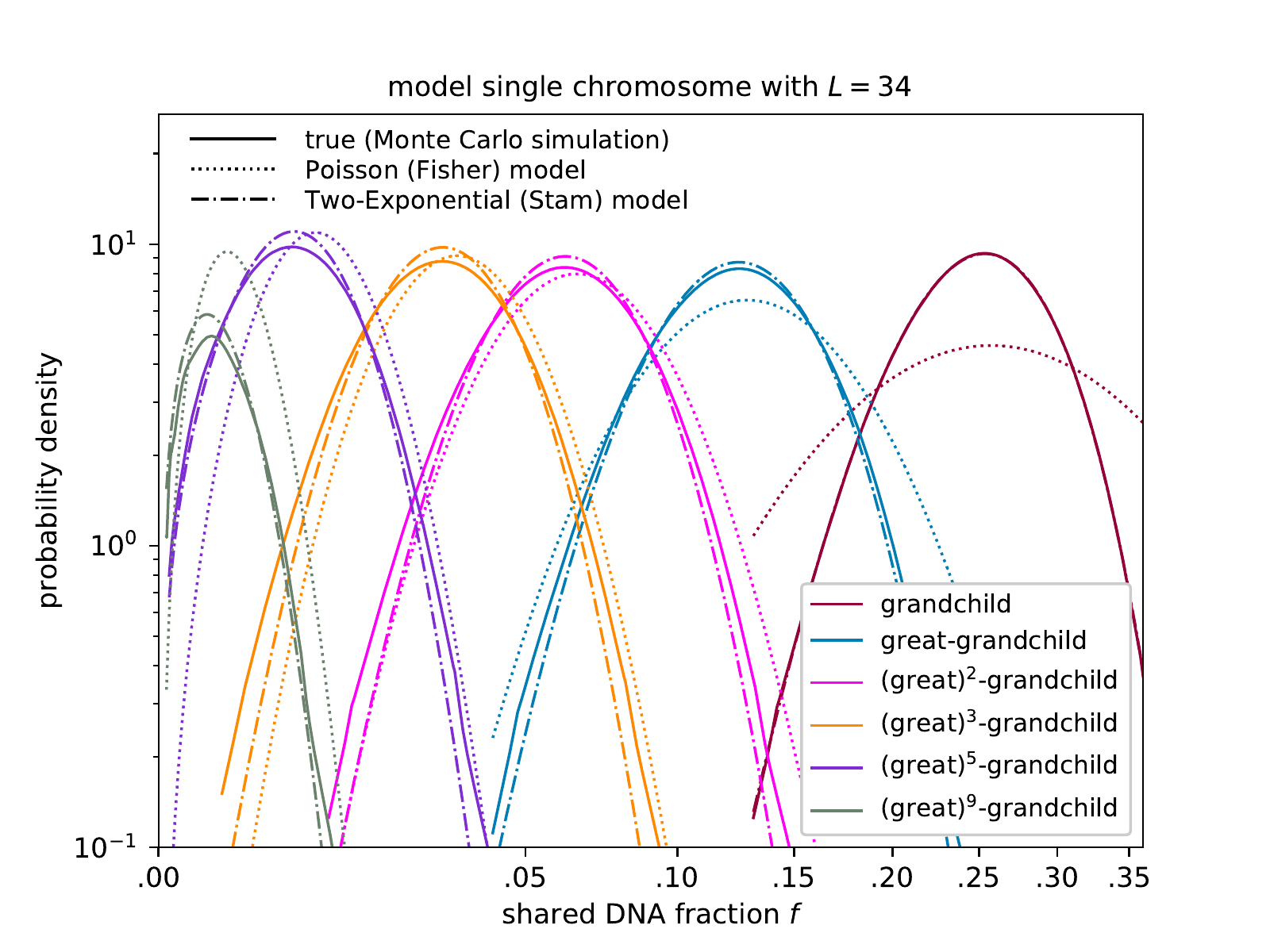}
\caption{Probability distribution of shared DNA fraction for lineal
  descendants, in a simplified model with a single chromosome of
  length 34 Morgans.  Solid curves are accurate simulations.  Dotted
  and dash-dot curves are respectively the Poisson and Two-Exponential
  models (see text).  The fraction $f$ is plotted on a square-root
  distorted scale as explained in \S \ref{normapprox}.  Probability density is
  plotted on a logarithmic scale, so that a normal distribution in the
  scaled abscissa would be exactly an inverted parabola.}
\label{fig1}
\end{figure}
This is a striking analytical result, especially the unexpected
occurrence of a modified Bessel function $I_1$ in equation
\eqref{eqpdf1}.  In \S\ref{fisherbennett} we give a derivation in
modern notation, also pointing out where additional assumptions are
made.  The key question is, how accurate is equation \eqref{eqpdf1}?
At the time it was derived, there was no way to know; but it is
straightforward now to compare it to exhaustive Monte Carlo computer
simulations (described in \S\ref{simulation}).  Figure \ref{fig1}
shows results for the conceptually simple case of a single chromosome
whose length is the same as the total human genome.  Apart from giving
the correct mean (by construction), equation \eqref{eqpdf1} agrees
poorly with accurate calculation. Its assumptions, which seemed
reasonable at the time, turn out to be unjustified.
For intuition about how the Poisson model fails, we can note that it
positions the starting positions of common segments randomly.  In
actuality, common segments are self-avoiding, because they are
separated by the segments common to other ancestors.  Self-avoidance
decreases the variance of their number, producing the narrower
``true'' distributions in Figure \ref{fig1}.

\subsection{Assuming Two Exponentials}

Understanding the self-avoidance issue, Stam \cite{stam}, in 1980,
modeled two alternating Poisson processes along a chromosome,
representing segments in common (state 1), or not in common (state 0),
with a specified ancestor.  For later use in \S\ref{fittwo}, we derive a slight
generalization.  Suppose that $(p_0,\;p_1=1-p_0)$ are the
probabilities of starting in states $(0,1)$ at the beginning of the
chromosome, and suppose that $(\lambda_0, \lambda_1)$ are the rate
constants for the two exponentials, so that the segment lengths $x$
are drawn from the exponential probability distributions
with densities $p(x|\lambda_i) =
\lambda_i \exp(-\lambda_i x)$, for $i=0,1$.  Then the resulting
probability density for the common fraction $f$, $0 < f < 1$, denoted
$P^*$, can be shown to be
\begin{equation}
\begin{split}
  P^*(f\,|p_0,&p_1,\lambda_0,\lambda_1) = e^{-(1-f)\lambda_0-f\lambda_1}\\
  &\times\left[\left(\sqrt{(1-f)/f}p_0+\sqrt{f/(1-f)}p_1\right)
  \sqrt{\lambda_0\lambda_1}
  I_1\left(2\sqrt{f(1-f)\lambda_0\lambda_1}\right)\right.\\
  &\left. + (\lambda_0p_0+\lambda_1p_1)I_0\left(2\sqrt{f(1-f)\lambda_0\lambda_1}\right)
  \right]
\end{split}
\label{eqmaster}
\end{equation}
with modified Bessel functions $I_0$ and $I_1$ (see \S\ref{derive2exp}
for derivation).  The massed probabilities that the chromosome is
entirely in state 0 or 1 are
\begin{equation}
  P_i^*(\text{all}|\,p_0,p_1,\lambda_0,\lambda_1)
  = p_i \exp(-\lambda_i),\qquad i=0,1
\label{eqmassed}
\end{equation}
The normalization is as expected,
\begin{equation}
  1 =P_0^*(\text{all}\,|p0,p1,\lambda_0,\lambda_1)
  +P_1^*(\text{all}\,|p0,p1,\lambda_0,\lambda_1)
  + \int_0^1 P^*(f|p_0,p_1,\lambda_0,\lambda_1) df 
\end{equation}

In this generalized setting, Stam's \cite{stam} results set $p_1 =
2^{-k}$, $p_0 = 1-2^{-k}$, $\lambda_1 = kL$, and $\lambda_0 =
kL/(2^k-1)$.  This ratio of the $\lambda$'s is set by the
requirement that the process spend $2^k-1$ times as long, on average,
in state 0 (the $2^k-1$ ancestors not of interest) as in state 1 (the
one ancestor of interest).  The result for fractional shared DNA after
$k$ meioses is thus expressed in terms of master equation \eqref{eqmaster}
as
\begin{equation}
p(f|k) = 2 P^*(2f|\,1-2^{-k},2^{-k},kL/(2^k-1),kL)
\label{eq2exps}
\end{equation}
with the factor of 2 on $f$ and $P^*$ to renormalize the denominator
convention from haploid to diploid chromosome total lengths (e.g., to
have grandchild share 1/4 with paternal grandparent instead of
the equivalent 1/2 of the paternal haploid).

The dash-dot curves in Figure \ref{fig1} show the accuracy of the
two-exponential model.  It is exact for grandchild (explained below,
see also \cite{walters}), and better than Fisher for other
descendants, but not close to exact.  In particular, it is seriously
in error for relationships that imply small, but nonzero, shared
fractions.  These are often the relationships most important to
forensic genealogists.

\begin{figure}[!tb]
\centering
\includegraphics[width=4.5in]{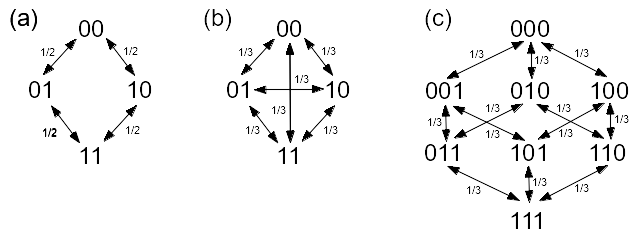}
\caption{Markov graphs for two and three meioses. (a) Two meioses
  (actual). (b) Two meioses (two-exponential model). (c) Three meioses
  (actual). The true return times to a specified ancestor (e.g., 00 or
  000) are those associated with hypercube graphs (a) and (c).  The
  two-exponential model approximates the hypercube by the complete graph,
  e.g., (b).}
\label{fig3}
\end{figure}

\subsection{Markov Models}
\label{markovmodels}

Donnelly \cite{donnelly} first elucidated the precise nature of the
failure of the two-exponential model (indeed, any exponential model)
to give exact results.  (See \cite{tiret} for more recent
references.) Consider the $k$ meioses present in a
descendant's chromosome as binary switches with values 0 or 1.  One
particular setting of these switches, call it the all-zeros value
$00\ldots0$, yields DNA in common with a specified ancestor.  Each
crossover junction on the chromosome flips exactly one switch
(changing exactly one zero to one or vice versa).  So, along the
length of the chromosome, the junctions collectively generate a Markov
chain, in particular a random walk on a hypercube graph (see Figure
\ref{fig3}).

At each vertex, the distribution of lengths before a transition is
$p(x)dx = k \exp(-k)$, so, in particular, this is the distribution of
the individual lengths of common segments, just as in the
two-exponential model.  However, the distribution of lengths while in
the state ``{\em not} $00\ldots0$'' is not exponential at all.
Rather, it is, in any one realization, the sum of $m-1$ such
exponential draws, that is a Gamma distribution of order $m-1$ where
$m$ is itself drawn from the integer distribution of {\em revisit
  times} to a vertex under random walks on the hypercube. The
two-exponential model in effect substitutes a complete graph like
Figure \ref{fig3}(b), whose revisit time is a binomial probability,
for the true hypercube Figure \ref{fig3}(a) or \ref{fig3}(c).  These
coincide only for $k=1$ (a single meoisis). This is the the grandchild
relationship where the two-exponential model was seen in Figure
\ref{fig1} to be exact.

\begin{figure}[!tb]
\centering
\includegraphics[width=5.5in]{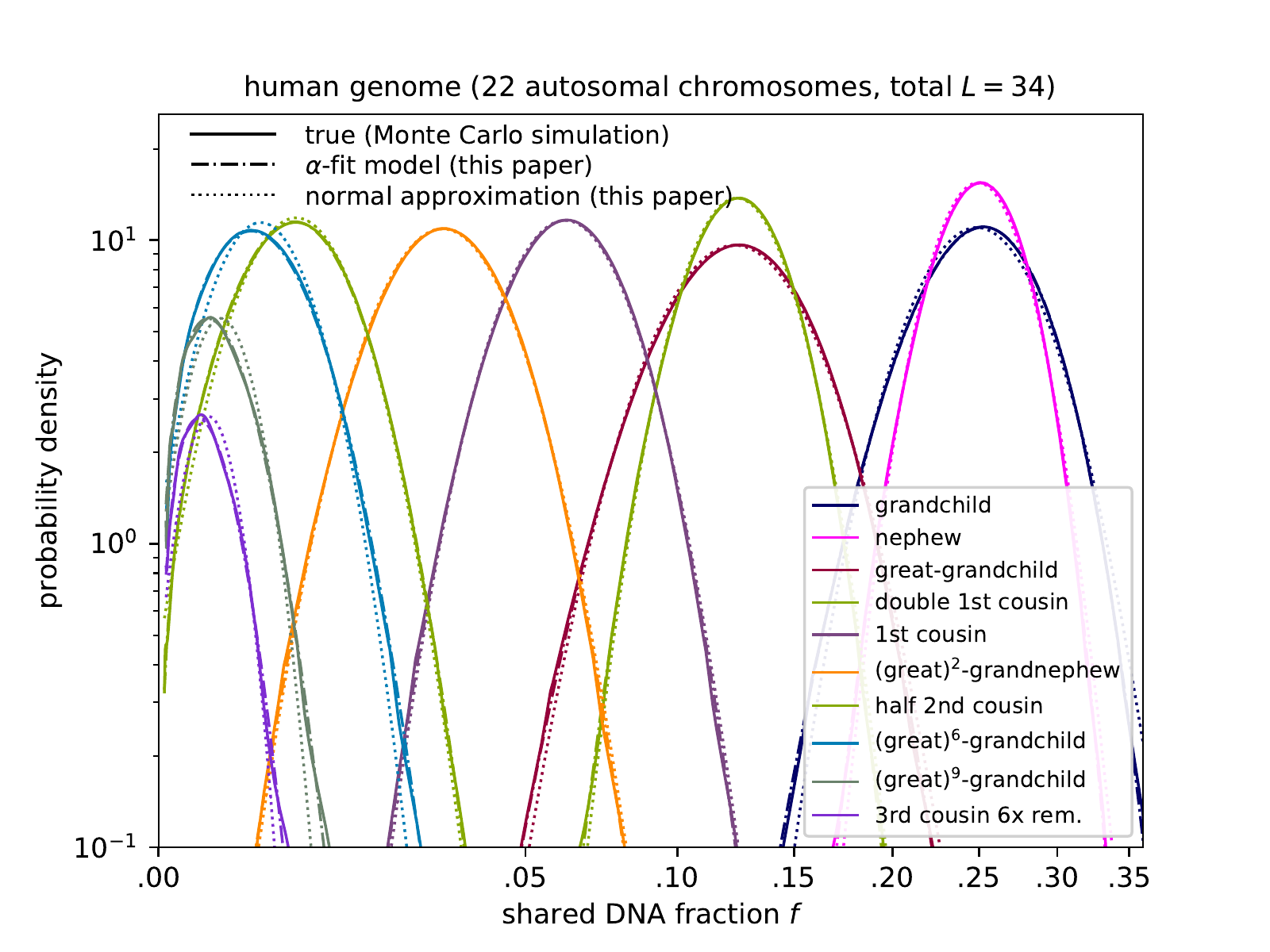}
\caption{Probability distribution of shared DNA fraction for various
  relationships.  The differences between the accurate Monte Carlo
  estimates (solid curve) and the model given in this paper (dash-dot
  curve) are only barely visible in the far tails of the distributions
  and amount to a few parts in a thousand as measured by
  Kolmogorov-Smirnov or Kullback-Liebler distances. The dotted curves
  are the normal approximation to the model in square-root
  coordinates (see \S\ref{normapprox}).}
\label{fig4}
\end{figure}

\section{Results}
\label{results}

The Poisson and two-exponential models are inadequate to our purposes
for three reasons.  First, they are not accurate enough.  Second, they
don't generalize to relationships other than lineal descendants.  For
example, Figure \ref{fig4} shows distributions of shared DNA fraction
(as determined by Monte Carlo) for a selection of other relationships.
One sees (as mentioned above) differences between grandchild and
nephew, or great-grandchild and double first cousin, that we need to
be able to model.  Third, the previous models don't point us towards
the construction of a multivariate model for use when shared DNA
fractions to multiple related individuals are known.  We now address
all three of these issues.

\subsection{Fitted Two-Exponential Model}
\label{fittwo}

One main result in this paper is a purely empirical (but genetically
motivated) fit to distributions such as those shown in Figure
\ref{fig4}.  We saw that the inaccuracy of Stam's two-exponential
approximation is because, on the hypercube with $2^k$ vertices, the
complement of a single vertex is not the complete graph with $2^k-1$
vertices.  But, we might ask, is there some fictitious number of
vertices different from $2^k-1$ that gives a better approximation?
That number need not even be an integer, because the intended use is
as a real-valued number in equation \eqref{eqmaster}.  As an example,
what if we imagine a graph in Figure \ref{fig3}(b) that has not 4
vertices, but 4.414, where the ratio $\alpha$ (here $4.414/4 = 1.104$)
is a fitted parameter. Making that change alone would alter the
desired mean fractions of $2^{-k}$, so we must also scale $f$ to
preserve its mean.  The specific proposal is to replace equation
\eqref{eq2exps} by the scaled equation
\begin{equation}
  p(f|k) = \frac{2}{\alpha} P^*\left[\frac{2f}{\alpha}\,\bigg|\, 1-\frac{1}{\alpha 2^{k_1}},
  \frac{1}{\alpha 2^{k_1}},\frac{k_2L}{\alpha 2^{k_1}-1},k_2L\right]
\label{mod2exp}
\end{equation}
(compare to equation \ref{eq2exps}), where $k_1$, $k_2$, and $\alpha$
are empirically determined for any particular relationship (e.g.,
second cousin three times removed) and chromosome or genome length
$L$.  The reason for splitting $k$ into the two parameters $k_1$
and $k_2$ is that, for relationships other then lineal descendants,
the ``effective'' number of meioses that determine the length of
a common segment ($k_2$) is different from the integer power of two that gives
the mean common fraction ($k_1$).  Specifically,
the best values for $k_2$ can be half-integers where changing the state
of one switch exposes another that will be right by chance half
the time.  Section \ref{cryuncle} works the example for
the case of uncle/nephew.

In the following, we fix the value $L=44$ (Morgans), and compare the
result to simulations of the autosomal human genome, 22 chromosomes
each having its observed genomic length.  One may think of the value
44 as being 34 Morgans (the sum of the genetic lengths of the
autosomal chromosomes) plus some fraction of a Morgan for each
inter-chromosomal decorrelation; but, in fact, the value is chosen
simply empirically as a single value that produces excellent fits to
across a wide range of relationships.  Our results are not sensitive
to the chosen value of $L$ (with correspondingly different values
$\alpha$).  We give results for the parameters in equation
\eqref{mod2exp} not only for the case of lineal descendants, but also
for other relationships.

We fit for the best value $\alpha$, by minimizing the
Kolmogorov-Smirnov \mbox{(K-S)} distance $D_{KS}$ between equation
\eqref{mod2exp} and the Monte Carlo simulations.  $D_{KS}$ is, by
definition, the maximum absolute difference between the two cumulative
distribution functions at any $f$.  Also of interest (discussion
below) is the Kullbach-Leibler divergence or relative entropy of the
two distributions, denoted $D_{KL}$.  Table \ref{tab1} gives the
values of $\alpha$ obtained, along with $D_{KS}$ and $D_{KL}$ for a
representative selection of relationships.  The quantity
$D^\prime_{KL}$ will be discussed in \S\ref{normapprox}
below. Supplemental Table S1 gives results for a much more complete
set of relationships, and has accuracy measures comparable to those in
Table \ref{tab1}.  It is reassuring that the fitted values for
$\alpha$ are not too different from 1, implying that the best-fitting
fictitious numbers of vertices for the complete graphs are not too different
from the actual numbers in the hypercube graphs. (You can get lost
in a hypercube---but not too lost.)

\begin{table}[ht]
  \centering
  \caption{Parameters $\alpha,k_1,k_2$ for the fitted model for representative relationships,
    and accuracy of the fits. }
  \begin{tabular}[t]{lccc|ccc}
    \toprule
    &\multicolumn{3}{c}{parameters}&\multicolumn{3}{c}{accuracy measures}\\
    &$\alpha$&$k_1$&$k_2$&$D_{KS}$&$D_{KL}$&$D^\prime_{KL}$\\
    \midrule
    grandchild&0.973&1&1&0.0010&0.0011&0.0051\\
    great-grandchild&1.036&2&2&0.0016&0.0011&0.0027\\
    (great)$^6$-grandchild&1.380&7&7&0.0027&0.0039&0.0293\\
    nephew&1.055&1&2.5&0.0011&0.0009&0.0029\\
    (great)$^2$-grandnephew&1.381&4&5.5&0.0030&0.0010&0.0014\\
    half-sibling&1.030&1&2&0.0005&0.0009&0.0031\\
    half-nephew&1.018&2&2.5&0.0011&0.0009&0.0022\\
    1st cousin&1.123&2&4&0.0015&0.0009&0.0016\\
    1st cousin 3$\times$ removed&1.398&5&7&0.0022&0.0005&0.0014\\
    double 1st cousin&1.321&1&3&0.0034&0.0011&0.0013\\
    half 1st cousin&1.231&3&4&0.0032&0.0011&0.0018\\
    half 2nd cousin 2$\times$ removed&1.435&7&8&0.0028&0.0052&0.0262\\
    3rd cousin 6$\times$ removed&1.378&12&14&0.0047&0.0126&0.0724\\
    \bottomrule
  \end{tabular}
  \label{tab1}
\end{table}

The typical accuracies obtained, a few parts in $10^3$, have these
interpretations: For $D_{KS}$, percentile points may be in error by at
most that amount, e.g., a 5\% critical region might actually be a
4.9\% or 5.1\% region.  For $D_{KL}$, the actual mean log
probability of observed data (per draw) may be parts in a thousand
greater than that indicated by the model.  These are negligible errors
for the intended application to forensic genealogy.  Figure
\ref{fig4}, in which the differences are only barely visible as
differences between the solid and dashed curves, reinforces this
point.  Note that the total normalization (area under curve) for the
shown example of 3rd cousin 6$\times$ removed is less than the other
curves.  The balance is made up by the massed probability of zero
shared DNA for this distant a relationship, as given by equation
\eqref{eqmassed}.

Child and sibling are special cases.  Child has $f=1/2$ with massed probability 1,
that is, one haploid in each diploid chromosome.  Siblings share DNA in both
diploid copies.  Purely empirically, we find that the parameters
\begin{equation}
p_\text{sib}(f)df = P^*(f|0.5,0.5,161.5,161.5)df
\end{equation}
give an excellent fit with $D_{KS}=0.0008$.  (Roughly speaking, the fact that siblings sum
two independent random variables---the two diploid copies---reduces the
variance if $f$, which is well fit by the larger effective value of
$kL$.)

\subsection{Normal Approximation}
\label{normapprox}

Not immediately apparent in equation \eqref{eqmaster} is, between the
exponential and Bessel function factors, where the peaked
distributions seen in Figure \ref{fig4} actually come from.  This
becomes clear if we replace the
Bessel functions by the two limiting cases of their arguments, either
$\gg 1$ or $\ll 1$. We find two possibilities for the leading
exponential terms
\begin{equation}
  P^*(f) \sim
  \begin{cases}
    \exp[-(\sqrt{(1-f)\lambda_0} -
      \sqrt{f\lambda_1})^2],\quad\text{or}\\
    \exp[-\lambda_0-(\lambda_1-\lambda_0)f]
  \end{cases}
\label{leadterms}
\end{equation}
As shown in \S\ref{asympforms} below, these are both approximately
normal distributions, not in the variable $f$, but in the variable
$s\equiv \sqrt{f}$.  Moreover, in the limit $\lambda_0 \ll \lambda_1$
(equation \ref{mod2exp} when $2^{k_1}\gg 1$), both forms in equation
\eqref{leadterms} imply the same variance (in the variable $s$),
$\sigma^2 = 1/(2\lambda_1)$.  This explains why the curves in Figure
\ref{fig4} are, by eye, close to parabolic (the normal distributions
shown as dotted curves), and why their widths change only slowly as
the modal fractions $f$ become small.  It is also the reason that we
plotted Figures \ref{fig1} and \ref{fig4} with a square-root scaled
abscissa.  We determine the actual parameters $(\mu,\sigma)$ of our
normal models not from these asymptotic expansions, but from the Monte
Carlo simulations.  For intuition about where the square-root
coordinate comes from, we can note that square root is the so-called
variance stabilizing transformation for the Poisson distribution; so
the Poisson approximation in Fisher's model may be making its presence
felt.

It would be attractive to use a normal approximation
(in coordinate $s$) to the various distributions, because this would
then naturally generalize to a multivariate normal model that can
include correlations.  Is this accurate enough?  One estimate
of this the Kullback-Leibler divergence between
the normal approximation and the true distribution, written as
\begin{equation}
  D^\prime_{KL} \equiv \int_0^1 \log \left( \frac{p(s)}{q(s)} \right)\, p(s) ds
  = \left<\log p(s) \right> - \left<\log q(s) \right>
\end{equation}
where angle brackets denote expectation over the true distribution
$p(s)ds$, and $q(s)$ denotes the normal approximation.  The
interpretation is that, for events sampled from the true distribution,
$D^\prime_{KL}$ (which is always positive) is, per observation, the
mean excess log probability of the observation over that calculated by
the normal model.  If this were large, then the normal model would, on
average, reject plausible observations and would thus not be useful.
However, the quantity $D^\prime_{KL}$ was given, for a selection of
relationships, as the last column of Table \ref{tab1}.  For all but
very distant relationships (e.g., (great)$^6$-grandchild or 3rd cousin
6$\times$ removed) it is on the order of parts per thousand.  This is
small enough to make a normal model viable, so we may with some
confidence turn to the multivariate case.  Figure \ref{fig4} plotted
as dotted curves the normal approximation, visibly less perfect than
the fitted two-exponential model, but nevertheless remarkably
good.

\section{Multivariate Normal Model}
\label{mvariate}

Suppose that we have some set ${T_i}$, $i=1,\ldots I$, of hypothesized
family trees, and that we are given $J$ measurements
$f_j, \, j=1,\ldots,J$ of the fractional DNA matches between pairs of
individuals specified on each tree.  We immediately convert to
square-root variables by $s_j \equiv \sqrt{f_j}$. Probability densities
$p_F(f)df$ and $p_S(s)ds$ are related by the law of transformation of
probabilities,
\begin{equation}
p_S(s|i) = 2s\, p_F(s^2|i)
\end{equation}
Here the conditioning on $i$ means simply ``for the relationship
specified for $s_j$ in $T_i$''.  We denote the normal approximation to
$p_S(s|i)$ by $q(s|i)$.

If the measurements $s_j$ were independent, we could write the
likelihood of each family tree as
\begin{equation}
L_i = \prod_j p_S(s_j|i) \approx  \prod_j q(s_j|i)
\end{equation}
The Bayes odds comparing two hypotheses (up to a choice
of prior, to be discussed in \S\ref{priordiscussion}) would the ratio of their $L_i$'s.

\begin{table}[ht]
  \centering
  \caption{Correlation coefficient for shared DNA fraction for
  selected pairs of relationships.}
  \begin{tabular}[t]{lll}
    \toprule
    $j_1\;(A\leftrightarrow B)$&$j_2\;(A\leftrightarrow C)$&$r_{j_1j_2}$\\
    $B$ is $A$'s:&$C$ is $B$'s:&\\
    \midrule
    sibling&child&0.76\\
    uncle&child&0.59\\
    half 3rd cousin&child&0.74\\
    (many relationships)&child&0.50--0.77\\    
    grandparent&sibling&0.73\\
    great-great grandparent&sibling&0.81\\
    sibling&grandchild&0.40\\
    uncle&grandchild&0.38\\
    (many relationships)&grandchild&0.37--0.55\\    
    sibling&great-grandchild&0.27\\
    \bottomrule
  \end{tabular}
  \label{tab2}
\end{table}

We turn to Monte Carlo simulations to see when or whether the
assumption of independence might be justified.  Table \ref{tab2} gives values
for the correlation coefficient
\begin{equation}
  r_{j_1j_2}
  = \frac{\left<(s_{j_1} - \left<s_{j_1}\right>)(s_{j_2} - \left<s_{j_2}\right>)\right>}
  {\left[\left<(s_{j_2} - \left<s_{j_2}\right>)^2\right>
    \left<(s_{j_2} - \left<s_{j_2}\right>)^2\right>\right]^{1/2}}
 = \frac{\text{cov}(s_{j_1},s_{j_2})}{\sigma(s_{j_1})\sigma(s_{j_2})}
\end{equation}
where angle brackets denote the mean of large numbers of trials.
Values for a large number of relationship pairs are given systematically in
Supplemental Table S2.  One finds substantial correlations, especially
when, for example, unknown person UP is compared both to a person $B$
and to a sibling, child, or grandchild of $B$.

A multivariate normal model that includes correlations must utilize the covariance
matrix
\begin{equation}
S_{j_1j_2} \equiv \text{cov}(s_{j_1},s_{j_2}) = r_{j_1j_2} \sigma_{j_1} \sigma_{j_2}
\label{eqdefcov}
\end{equation}
(with diagonal elements $S_{jj} = \sigma^2_j$).  If $\Sbf^{-1}$ denotes the matrix
inverse of $\Sbf$, then one computes from the measured values $\widehat{s_j}$
\begin{equation}
  \chi^2 = \sum_{j_1,j_2} (\widehat{s}_{j_1} - \left<s_{j_1}\right>) S^{-1}_{j_1j_2}
  (\widehat{s}_{j_2} - \left<s_{j_2}\right>)
\label{chisqeq}
\end{equation}
in terms of which the multivariate normal log likelihood is
\begin{equation}
\log L_i = -\half \chi^2 - \half J\log(2\pi) -\half \log\det\Sbf_i
\label{likelihoodeq}
\end{equation}
We have appended an index $i$ to $\Sbf$ to underscore the fact that the covariance
matrix depends on the hypothesis $T_i$, not on the measured values $\{\widehat{s}_j\}$.
Bayes model comparison requires not just the likelihood $L_i$, but also prior
probabilities $p_{0i}$.  We discuss these in the next section.

\subsection{Practical Considerations}
\label{priordiscussion}

\begin{figure}[!tb]
\centering
\includegraphics[width=5.0in]{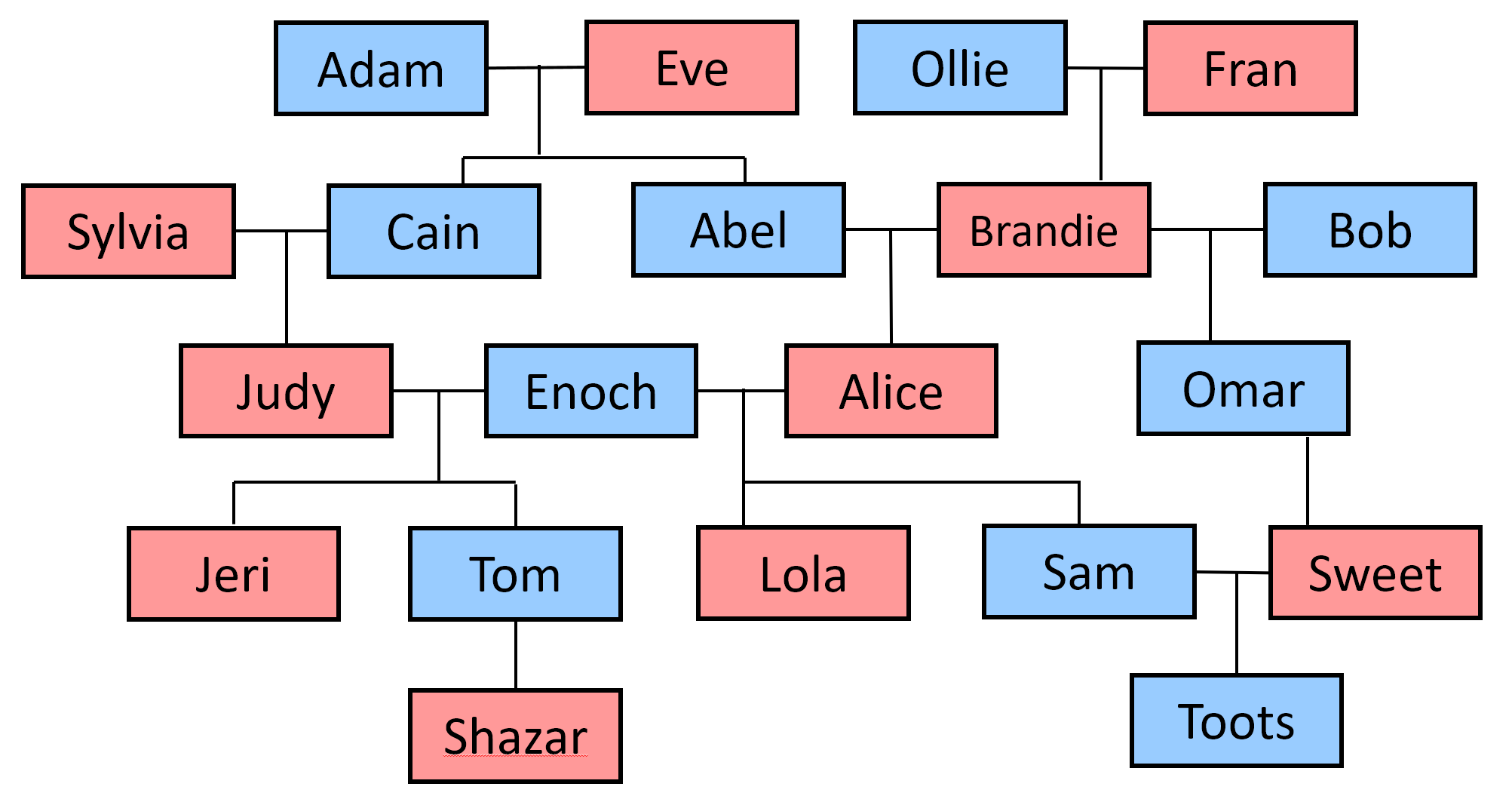}
\caption{A fictitious, but possible, family tree with endogeny,
  half- and double-relationships.}
\label{fig5}
\end{figure}

The methods developed in this paper have been used to construct a
computer code for use by forensic genealogists.  We describe here some
of the practical issues that arise.

In practice, genealogists cannot be confident that their set of stated
hypotheses $\{T_i\}$ include all viable possibilities.  In that case,
a Bayesian calculation will select the most probable hypothesis, even
if that hypothesis is extremely unlikely.  This, if undetected, is
undesirable.  Needed in practice is a rejection criterion for unlikely
models, notwithstanding that ``rejection'' is usually thought of as a
frequentist, not Bayesian, concept.  If synthetic sets of measured
values $\{\widehat{s}_j\}$ are drawn from the multivariate normal model of
hypothesis $T_i$, then the values $\chi^2$ (equation \ref{chisqeq})
will be chi-square distributed with $J$ degrees of freedom.  This
provides a one-tail rejection criterion for any chosen critical value
(tail probability) $p_c$.  In practice, because of various unmodeled
effects in the measured data, values of $p_c$ as large as $0.05$ or
$0.01$ may often reject true hypotheses; a value $p_c = 0.001$ is
found to be useful.

The Bayes odds (ratio of probabilities) for two hypotheses $i_1$ and $i_2$
that survive rejection is
\begin{equation}
\text{Odds}_{\,i_1i_2} = \frac{P_{i_1}}{P_{i_2}} = \frac{L_{i_1}}{L_{i_2}} \times \frac{p_{0i_1}}{p_{0i_2}} 
\label{oddseq}
\end{equation}
where the $L_i$'s are likelihoods (equation \ref{likelihoodeq}), the
$p_{0i}$'s priors.  It is well understood in the statistics literature
that, in the context of model selection, one cannot simply take
$p_{0i} = \text{constant}$ to be the non-informative priors of choice
(see, e.g., discussions in \cite{berger} and \cite{wasserman}).  A
part of equation \eqref{likelihoodeq} depends on the structure of the
model $T_i$, that is, its number of parameters, $J$, and the
narrowness of their probable ranges, encoded as the value of the
determinant $\log\det\Sbf_i$.  With constant priors, Bayes favors few
parameters with narrow ranges, the so-called {\em Bayes complexity
  penalty} or {\em Ockham factor} \cite{loredo}.  In some
applications, this is arguably is a desired feature---but not here. We
want to choose priors that will not strongly bias the selection of one
hypothesis over another when both are good fits (in the frequentist
sense) to the data.  Noting that in expectation
$\left<\chi^2\right> = J$, we choose
\begin{equation}
\log p_{0i} = \half \log\det\Sbf_i + \half (1+\log 2\pi)J_i
\end{equation}
This depends on the model structure only (not the data), so is legitimate
as a prior. Equation \eqref{oddseq} now becomes
\begin{equation}
  \text{Odds}_{\,i_1i_2} = \frac{P_{i_1}}{P_{i_2}} =
  \exp[-\half(\chi^2_{i_1} - \chi^2_{i_2})]\, \exp[+\half(J_{i_1} - J_{i_2})]
\label{newoddseq}
\end{equation}
which has the desired property. Almost always we will have $J_{i_1} =
J_{i_2}$, in which case equation \eqref{newoddseq} involves only the
$\chi^2$'s.

If the $T_i$'s are a complete set of (non-negligible)
hypotheses, as required by Bayes Theorem, then their absolute
individual probabilities are
\begin{equation}
P(T_i|\{\widehat{s}_j\}) = \frac{\text{Odds}_{\,ik}}{\sum_i\text{Odds}_{\,ik}}
\end{equation}
which is numerically the same for any choice of $T_k$ against which to
measure.

Although the Supplemental Material includes the extension of Tables 1
and 2 to many relationships (and, for correlation, many pairs of
relationships), it is not practical to catalog all possible
relationships that may arise.  Family situations that ultimately give
rise to forensic work are often complicated.  As a fictitious, but
possible, example, Figure \ref{fig5} shows a family tree with endogeny
(inbreeding), half, and double relationships.  In the figure, Shazar
and Toots are simultaneously third cousins and half first cousins.
Lola is Toots' aunt by descent from Adam and Eve, but Lola and Toots
are also both also descended from Ollie and Fran in a complicated way,
through half-siblings Alice and Omar.  A practical code must therefore
have the capability of doing new simulations, capturing the parameters
of the multivariate normal model, that is, the means $\left<s_{j}\right>$
and covariances $\text{cov}(s_{j_1},s_{j_2})$.  Our simulation is
described below in  \S\ref{simulation}.

One might ask: If we are going to do simulations anyway, then why do
we need a model (and the whole apparatus of this paper) anyway?  The
answer is that, in the multivariate case of $J>1$, there is no simple
or generally accepted model-independent way to estimate the
likelihoods $L_i$ from the simulation data.  Even with many trials, a
simulation does not densely populate the $J$-dimensional space of
measured values.  A model simultaneously estimates the local density
and interpolates among scattered points in high-dimensional space.

\section{Discussion}
Using a Monte Carlo simulation (\S\ref{simulation}), we studied the
probability distributions of fractional common DNA between human
individuals related in various ways---not just lineal descendants.  We
showed why previous analytic attempts at calculating these
distributions produced inaccurate results.  We then showed that a
particular functional form, defined by equations \eqref{eqmaster} and
\eqref{mod2exp} with essentially a single continuous fitted parameter,
reproduces the simulation data with accuracy of parts per thousand
(measured in multiple ways) over a wide range of complex relationships.

Analysis of that functional form, showed that it in turn can be well
approximated by a normal distribution, not in shared DNA fraction $f$,
but rather in $\sqrt{f}$.  That led us to propose a multivariate
normal model in $\sqrt{f}$ as a general framework for use in forensic
genealogy, one that evaluates the respective likelihoods of different
hypothesized family trees, given a set of shared DNA measurements.  We
gave a detailed description of that model in \S\ref{mvariate}.  These
ideas are being used in a code that we are developing in cooperation
with, and for use by, practicing forensic genealogists.

Centimorgan values for shared DNA as reported by genome database
organizations such as GEDmatch \cite{GEDmatch} are in no sense exact.
Not only are sequencing or other errors possible in the uploaded data,
but there are also algorithmic choices in the match calculation that
are not visible to the database user.  For this reason, and also
because the Morgan crossover model is itself not exact, it is found
useful to add a small constant measurement-error term $\epsilon^2$ to
the diagonal elements of the covariance matrix $\Sbf$ (equation
\ref{eqdefcov}).  In the absence of other information, the constant
$\epsilon = 0.03$ seems to work well, but the value should be
adjustable by the user.  In any particular instance, $\epsilon$ can be
adjusted to make known-true family trees give reasonable $\chi^2$
probabilities.  Since $\epsilon^2$ is added to $\Sbf$ in square-root
coordinates, its centimorgan equivalent varies with measured value.
For a constant $\epsilon=0.03$ the implied error is about $\pm 35$ at
100 centimorgans, rising to about $\pm 100$ at 1000 centimorgans.

\section{Methods and Materials}

\subsection{Poisson Model of Fisher and Bennett}
\label{fisherbennett}

Fisher \cite{fuller} and Bennett \cite{bennett} argued as follows:
After $k$ meioses of a haploid chromosome of length $L$ (in Morgans),
the mean length in common with any one ancestor is $2^{-k}L$. Further,
there are then on average $kL$ junction points from one or another of
the meioses, distributed uniformly randomly. An IBD segment from any
one ancestor terminates when it encounters any one such junction, so
its length is exponentially distributed, $p(x)dx=k\,\exp(-k)dx$.
Thus, for any specific number $n$ of such segments in common, their
total length $x$ is the sum of $n$ such exponential deviates.  Such a
sum is $\Gam(n,k)$ distributed.  We {\em assume} negligible
correlation between the mean number of common segments $\nbar$ and
their mean length $1/k$. The intuition for this is that the events
that start a segment (responsible for their number) are different from
the events that terminate a segment (responsible for their
lengths). The total length is then the product of the mean number and
mean length, implying
\begin{equation}
  \nbar = k 2^{-k} L
\end{equation}

Now further {\em assume} that the actual number $n$ of common segments
is Poisson distributed around its mean because (intuitively) the separated
segments occur independently of each other.  Then the distribution of
total common length $x$ is
\begin{equation}
\begin{split}
  p(x|k,\nbar)dx &= \sum_{n=1}^\infty \Gam(x|n,k)\Poi(n|\nbar)dx\\
  &= \exp(-\nbar-k x)\sqrt{k \nbar/x}\,I_1(2\sqrt{k \nbar x}) dx
\label{eqsum1}
\end{split}
\end{equation}
where $\Poi(k|\lambda)$ and $\Gam(x|n,\lambda)$ denote the Poisson
and Gamma distribution densities,
\begin{equation}
\Poi(k|\lambda) = \frac{\lambda^k}{k!}\exp(-\lambda),\quad
\Gam(x|n,\lambda) = \frac{\lambda^n x^{n-1}}{(n-1)!}\exp(-\lambda x)
\end{equation}
Remarkably, the sum yielding a modified Bessel function $I_1$ can be
done by Mathematica. The missing term in the sum, $n=0$, is the massed
probability of having no common segments, $p_0 = \Poi(0|\nbar)$.

Transforming the result from the variable $x$ to the shared DNA
fraction $f = x/(2L)$ gives the probability density for $f$ in the
range $0<f\le 1/2$,
\begin{equation}
  p(f|k,L)\,df = \exp[-k L(2^{-k}+2f)]\,kL\sqrt{2^{-k+1}/f}\,
  I_1\left(2kL\sqrt{2^{-k+1}f}\right)\,df
\label{eqpdf1aa}
\end{equation}
The reason for the factor $1/2$ in the definition of $f$
is that $f$ is here defined as
the fraction of total genome, including two diploid chromosomes, one of which
comes from an unrelated parent.  For the same reason,
we have $p(f|k,L)=0$ for $1/2 < f < 1$.  The
massed probability at $f=0$ is
\begin{equation}
p(0|k,L) = \exp(-k 2^{-k} L)
\end{equation}
Equation \eqref{eqpdf1} has the desired probability normalization,
\begin{equation}
1 = p(0|k,L) + \int_{f=0}^1 p(f|k,L) df
\end{equation}
The case $k=1$ corresponds to the relation of grandparent to
grandchild (separated by a single meiosis in the intervening parent);
$k=2$ corresponds to great-grandchild, etc.

Figure \ref{fig1} showed that the above argument, while seductive,
does not produce accurate results.

\subsection{Computer Simulation}
\label{simulation}
We take the total genetic length of the haploid human genome to be
3400 centimorgans, divided among 22 chromosomes in proportion to the
empirical lengths published by Family Tree DNA \cite{FTDNA,cM}.  We
discretize in segments of 0.1 centimorgan, each thus having
independent probability 0.001 of being a crossover location.  For a
single trial, full diploid genomes of every individual at the ``top''
of a given family tree (that is, not the result of a mating within the
tree) are initialized with an identifier unique to that individual.
Descendant genomes are then filled in by simulating random crossovers,
copying the unique identifier that is inherited by descent. For all
pairs of interest, a value for the IBD common fraction is summed by
direct comparison and saved as one draw from the desired
distribution. Trials are repeated typically $10^5$--$10^6$ times to
populate the distribution.

\begin{figure}[!tb]
\centering
\includegraphics[width=5.5in]{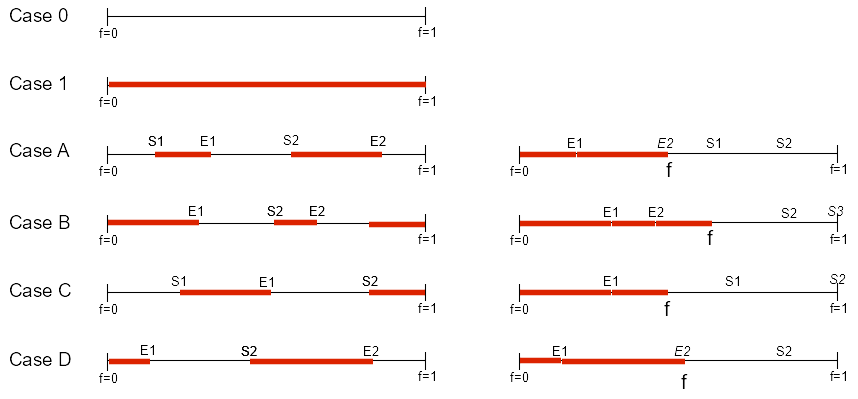}
\caption{ Cases for computing the probability of obtaining a total
  fraction $f$ for one of two alternating Poisson processes, states 0
  and 1.  State 1 segments are denoted as thicker red lines.
  In the right-hand column these have been grouped to the left,
  showing their total fraction $f$.  See text
  for details.}
\label{fig2}
\end{figure}

\subsection{Derivation of Equation \eqref{eqmaster}, the Two-Exponential Model}
\label{derive2exp}

Although the general idea is implicit in \cite{stam}, we give here
a streamlined and slightly generalized derivation.  There are six
disjoint possibilities to consider.  We denote them by (0,1,A,B,C,D),
as shown in Figure \ref{fig2}. Cases 0 and 1 represent the massed probability
of staying in state 0 or 1 for the entire length.  The respective
probabilities are
\begin{equation}
  P_i^*(\text{all}|\,p_0,p_1,\lambda_0,\lambda_1)
  = p_i \Poi(0,\lambda_i) = p_i \exp(-\lambda_i),\qquad i=0,1
\end{equation}
where $\Poi(j,\lambda)$ denotes the probability of drawing j from a
Poisson distribution with mean $\lambda$.

Cases A, B, C, and D, respresent the four possibilities of starting
and/or ending in states 0 or 1.  The figure shows the
correspondence. Red segments are state 1, with start and end positions
denoted by $S_k$ and $E_k$.  For each of A,B,C,D, we can imagine
regrouping all the segments in state 1 to the left, all in state 0 to
the right, as shown in the right column of the figure.  The dividing
line is the fraction $f$ of the length in state 1.  Now notice that
every $S_k$ and $E_k$ is either an ``interior'' junction, which can be
anywhere in its respective state, or a ``pinned'' junction that is
located exactly at $f$ (state 1) or $1$ (state 0).  This is enough to
immediately write the probability that $f$ lies between $f$ and $f+df$
in the four disjoint cases as the product of a starting probability
($p_0$ or $p_1$), two interior Poisson probabilities for the observed
numbers of interior junctions, and a pinned probability ($\lambda_0df$
or $\lambda_1df$), of course summed over all possible numbers of
segments:
\begin{equation}
\begin{split}
  P(f\cap A) df &= p_0\sum_{i=1}^\infty \Poi(i-1|f\lambda_1)
  \Poi(i|(1-f)\lambda_0)(\lambda_1 df)\\
  P(f\cap B) df &= p_1\sum_{i=1}^\infty \Poi(i|(f\lambda_1)
  \Poi(i-1|(1-f)\lambda_0)(\lambda_0 df)\\
  P(f\cap C) df &= p_0\sum_{i=0}^\infty \Poi(i|f\lambda_1)
  \Poi(i|(1-f)\lambda_0)(\lambda_0 df)\\
  P(f\cap D) df &= p_1\sum_{i=0}^\infty \Poi(i|f\lambda_1)
  \Poi(i|(1-f)\lambda_0)(\lambda_1 df)
\end{split}
\end{equation}
The sums can all be done analytically (in Mathematica), giving
the disjoint probabilities
\begin{equation}
\begin{split}
  P(f\cap A) df &=  p_0 e^{-(1-f)\lambda_0-f\lambda_1}
  \sqrt{(1-f)/f}\sqrt{\lambda_0\lambda_1}
  I_1\left(2\sqrt{f(1-f)\lambda_0\lambda_1}\right)\\
  P(f\cap B) df &= p_1 e^{-(1-f)\lambda_0-f\lambda_1}
  \sqrt{f/(1-f)}\sqrt{\lambda_0\lambda_1}
  I_1\left(2\sqrt{f(1-f)\lambda_0\lambda_1}\right)\\
  P(f\cap C) df &=  p_0 e^{-(1-f)\lambda_0-f\lambda_1}
  \lambda_0
  I_0\left(2\sqrt{f(1-f)\lambda_0\lambda_1}\right)\\
  P(f\cap D) df &= p_1 e^{-(1-f)\lambda_0-f\lambda_1}
  \lambda_1
  I_0\left(2\sqrt{f(1-f)\lambda_0\lambda_1}\right)
\end{split}
\end{equation}
The sum of the four cases gives equation \eqref{eqmaster}.

\subsection{Uncle/Nephew Is Best Fit by Half-Integral Number of Meioses}
\label{cryuncle}

In Table \ref{tab1}, for the relationships nephew or half-nephew and
descendants, recommended  values for the parameter $k_2$ are
half-integral.  We here give a heuristic explanation, with the case of
uncle/nephew as an example.  We denote the relevant individuals and
one of their diploid chromosomes as follows: Grandpa $(A)(A^\prime)$,
Grandma $(B)(B^\prime)$; their two children Uncle
$(A,A^\prime)(B,B^\prime)$ and Father $(A,A^\prime)(B,B^\prime)$; an
exogamous Mother $(\cdot)(\cdot)$; and Nephew, the son of Father and
Mother $(A,A^\prime,B,B^\prime)(\cdot)$.  We refer to the diploid
chromosomes in the order given as ``left'' and ``right''.

At a particular location on the chromosome of interest, there are five
relevant ``switches'' with values $0$ or $1$: 1. Is Nephew's left an
$(A,A^\prime)$ (0) or a $(B,B^\prime)$ (1)?  2. Is Uncle's left an $A$ (0) or an $A^\prime$
(1)?  3.  Is Father's left an $A$ (0) or an $A^\prime$ (1)? 4. Is
Uncle's right a $B$ (0) or a $B^\prime$ (1)? 5. Is Father's right a
$B$ (0) or a $B^\prime$ (1)?  Four patterns of these switches (in the
order 1 to 5) produce a match between Nephew and Uncle:
$000\cdot\cdot$, $011\cdot\cdot$, $1\cdot\cdot 00$, $1\cdot\cdot 1 1$.
Here ``dot'' means either value, 0 or 1.  Four other patterns produce no
match: $001\cdot\cdot$, $010\cdot\cdot$, $1\cdot\cdot 01$, and
$1\cdot\cdot 10$.  These eight patterns exhaust all possibilities.
Since half of them produce a match, we must have $k_1 = 1$ in equation
\eqref{mod2exp}.

Now for $k_2$, consider just the three ``active'' switches in each of
the eight patterns, i.e., those not denoted by a dot.  If changing any
one switch terminated a match between Nephew and Uncle, then we would
have $k_2 = 3$, by the argument in \S\ref{markovmodels}.  But, here,
if we change the first switch, then the dots change positions (from
2-3 to 4-5 or vice versa), exposing two different non-dotted
positions.  Half the time, the cases $00$ or $11$, the newly exposed
positions will happen to continue a match.  So switch number 1 is
really only half a switch and the heuristicy trial value is $k_2
= 2.5$.  The validation is that with this value and a fitted value for
$\alpha$, equation \eqref{mod2exp} gives an excellent fit, as seen
in Figure \ref{fig4}.

\subsection{Asymptotic Forms Suggest Square-Root Fraction Coordinate}
\label{asympforms}

The argument of both Bessel functions in equation \eqref{eqmaster} is
\begin{equation}
  \eta = 2\sqrt{f(1-f)\lambda_0\lambda_1}
\end{equation}
For $\eta \gg 1$,
\begin{equation}
I_0(\eta) \approx I_1(\eta) \approx \frac{1}{\sqrt{2\pi}\eta}\exp(\eta)
\end{equation}
and equation \eqref{eqmaster} can be written
\begin{equation}
  P^*(f) = (\text{sub-exponential factors})\times
  \exp[-(\sqrt{(1-f)\lambda_0} - \sqrt{f\lambda_1})^2]
\label{suggestive}
\end{equation}
Since we are generally interested in $f \ll 1/2$,
equation \eqref{suggestive} is suggestive of a normal distribution
in the coordinate $s\equiv\sqrt{f}$.  One easily calculates that
the maximum exponential argument occurs at $s_\text{max} =
\sqrt{\lambda_0}/(\sqrt{\lambda_0} +\sqrt{\lambda_1})$ and that
the second derivative there implies a variance,
\begin{equation}
\sigma^2 = \half \frac{\lambda_1}{(\lambda_0+\lambda_1)^2}
\approx \half \frac{1}{\lambda_1}
\end{equation}
where the last approximation is for $\lambda_0 \ll \lambda_1$ (compare
equations \ref{eq2exps} and \ref{mod2exp}).  For the opposite asymptotic limit
$\eta \ll 1$, we have $I_0(\eta) \approx 1$ and $I_1(\eta)\approx
  \half \eta$.  Now
\begin{equation}
  P^*(f) = (\text{sub-exponential factors})\times
  \exp[-\lambda_0-(\lambda_1-\lambda_0)s^2]
\label{suggestive2}
\end{equation}
implying a normal distribution in $s$ with mean zero and
\begin{equation}
\sigma^2 = \half \frac{1}{(\lambda_1-\lambda_0)}
\approx \half \frac{1}{\lambda_1}
\end{equation}
The fact that both asyptotic limits give (for $\lambda_0 \ll
\lambda_1$) the same variance in the $s$-coordinate, even as the means
go to zero, explains the utility of the square-root distorted scale in
Figures \ref{fig1} and \ref{fig4}.

\subsection*{Acknowledgments}
We have benefitted from discussions with Margaret Press and Colleen
Fitzpatrick, founders of the DNA Doe Project \cite{dnadoe}.

\section*{Supplemental Materials}

The Supplemental Materials can be found on GitHub at\\
https://github.com/whpress/forensicgenealogy and consist of computer files, some large, as follows:

\begin{itemize}

\item File {\em SupplementalTable1.tsv} (7KB) Values $k_1$, $k_2$, $\alpha$,
$D_{KS}$, $D_{KL}$, $D^\prime_{KL}$ for the fitted models of 96
relationships in the files {\em dependsdata\_*\_23.txt}.  Table 1 in
the main text gave a representative selection.

\item File {\em SupplementalTable2.xlsx} (17 KB):  Coefficients of correlation
$r$ for 173 pairs of relationships drawn from {\em bigrun\_data.out}, above.
Table 2 in the main text gave a representative selection.

\item Files {\em dependsdata\_*\_23.txt} (each about 50 MB):
  Centimorgans of common DNA for one million Monte Carlo simulations
  of the relationships (and also descendants $\le 9$ times removed)
  for lineal descendant, 1st cousin, 2nd cousin, 3rd cousin, double
  1st cousin, nephew, half 1st cousin, half 2nd cousin, half third
  cousin, and half nephew.

\item File {\em bigrun\_data.out} (470 MB): Centimorgans of common DNA for
one million Monte Carlo simulations among all pairs of descendants to
five generations of siblings and half-siblings.  Stored as zip file broken
into two pieces.  Can be reconstituted with 7-zip (e.g.).

\end{itemize}

\end{document}